\documentclass{aa}
\usepackage{natbib}
\bibpunct{(}{)}{;}{a}{}{,}
\usepackage[dvips]{graphics}
\usepackage[psamsfonts]{euscript}
\usepackage[psamsfonts]{amssymb}
\usepackage{amsmath}

\newcommand{\nc}   {\newcommand}
\nc{\dst}   {\displaystyle}
\nc{\dsfrac}[2]{{\dst\frac{#1}{#2}}}
\nc{\cm}    {\mathrm{cm}}
\nc{\gcc}   {{\rm gm}\:{\rm cm}^{-3}}
\nc{\K}     {\mathrm{K}}
\nc{\kms}   {{\rm km}\:{\rm s}^{-1}}
\nc{\CC}    {\EuScript{C}}
\nc{\FK}    {F_\mathrm{K}}
\nc{\Fh}    {F_\mathrm{h}}
\nc{\Em}    {\tilde{E}}
\nc{\Ema}   {\tilde{E}_1}
\nc{\Emn}   {\tilde{E}_N}
\nc{\Etr}   {\tilde{E}_\mathrm{t}}
\nc{\Etra}  {\tilde{E}_{\mathrm{t}1}}
\nc{\Ein}   {\tilde{E}_\mathrm{in}}
\nc{\Eina}  {\tilde{E}_{\mathrm{in}1}}
\nc{\ER}    {E_\mathrm{R}}
\nc{\FR}    {F_\mathrm{R}}
\nc{\FLyc}  {F_\mathrm{Lyc}}
\nc{\FKa}   {F_{\mathrm{K}1}}
\nc{\Fha}   {F_{\mathrm{h}1}}
\nc{\FKn}   {F_{\mathrm{K}N}}
\nc{\Fhn}   {F_{\mathrm{h}N}}
\nc{\FRa}   {F_{\mathrm{R}1}}
\nc{\FRn}   {F_{\mathrm{R}N}}
\nc{\Lya}   {Ly$\alpha$}
\nc{\Lyb}   {Ly$\beta$}
\nc{\Lyg}   {Ly$\gamma$}
\nc{\Ha}    {H$\alpha$}
\nc{\Hb}    {H$\beta$}
\nc{\Pa}    {Pa$\alpha$}
\nc{\PR}    {P_\mathrm{R}}
\nc{\Pg}    {P_\mathrm{g}}
\nc{\Pga}   {P_{\mathrm{g}1}}
\nc{\Pgn}   {P_{\mathrm{g}N}}
\nc{\Ta}    {T_a}
\nc{\Te}    {T_{\rm e}}
\nc{\Teff}  {T_{\rm eff}}
\nc{\nH}    {n_\mathrm{H}}
\nc{\nel}   {n_\mathrm{e}}
\nc{\xH}    {x_\mathrm{H}}
\nc{\rholim}{\rho_\mathrm{lim}}
\nc{\etalim}{\rholim/\rho_1}

\begin{document}

\title{The structure of radiative shock waves}
\subtitle{III. The model grid for partially ionized hydrogen gas}
\author{Yu.A.~Fadeyev\inst{1} \and D.~Gillet\inst{2}}
\offprints{D. Gillet, \email{gillet@obs-hp.fr}}
\institute{Institute for Astronomy of the Russian Academy of Sciences,
 Pyatnitskaya 48, 109017 Moscow, Russia
\and
 Observatoire de Haute-Provence - CNRS, F-04870
 Saint-Michel l'Observatoire, France}
\date{Received September 2000 / Accepted }

\abstract{The grid of the models of radiative shock waves propagating through
partially ionized hydrogen gas with temperature
$3000\K\le T_1\le 8000\K$ and density $10^{-12}~\gcc\le\rho_1\le 10^{-9}~\gcc$
is computed for shock velocities $20~\kms\le U_1\le 90~\kms$.
The fraction of the total energy of the shock wave irreversibly lost
due to radiation flux ranges from 0.3 to 0.8 for $20~\kms\le U_1\le 70~\kms$.
The postshock gas is compressed mostly due to radiative cooling in
the hydrogen recombination zone and final compression ratios are within
$1 <\rho_N/\rho_1\lesssim 10^2$, depending mostly on the shock velocity $U_1$.
The preshock gas temperature affects the shock wave structure due to the
equilibrium ionization of the unperturbed hydrogen gas,
since the rates of postshock relaxation processes are very sensitive
to the number density of hydrogen ions ahead the discontinuous jump.
Both the increase of the preshock gas temperature and the decrease of the
preshock gas density lead to lower postshock compression ratios.
The width of the shock wave decreases with increasing upstream velocity
while the postshock gas is still partially ionized and increases
as soon as the hydrogen is fully ionized.
All shock wave models exhibit stronger upstream radiation flux
emerging from the preshock outer boundary in comparison with
downstream radiation flux emerging in the opposite direction from the
postshock outer boundary.
The difference between these fluxes depends on the shock velocity
and ranges from 1\% to 16\% for $20~\kms\le U_1\le 60~\kms$.
The monochromatic radiation flux transported in hydrogen lines
significantly exceeds the flux of the background continuum and all
shock wave models demonstrate the hydrogen lines in emission.
}
\titlerunning{The structure of radiative shock waves. III}
\maketitle

\section{Introduction}

Main indicators of shock waves in atmospheres of radially pulsating stars
are enhanced hydrogen emission, double profiles of absorption lines
and discontinuities in radial velocity curves.
Most prominently these features are observed in
W~Vir \citep{Lebre:1992},
RV~Tau \citep{Gillet:1989}
and Mira type \citep{Alvarez:2000}
pulsating variables.
The ratio of the atmosphere scale height to the stellar radius in these
stars is $H/R\gtrsim 10^{-2}$ and the linear theory of stellar pulsation
admits the existence of progressive waves propagating through
the stellar atmosphere \citep{Unno:1965,Pijpers:1993}.
The running waves must inevitably transform into shock waves
due to nonlinear effects and
this conclusion is corroborated by hydrodynamic calculations of radial
pulsations of low--mass late--type giants
\citep{Wood:1974,Fadeyev:1981,Bowen:1988,Fadeyev:1990}.

Though the role of shock waves as the principal mechanism of mass loss from
late--type stars is still disputable \citep{Hoefner:1997}
it is nevertheless undoubted that shock--driven gas flows are responsible 
for formation of circumstellar shells.
It must be noted that the postshock velocity near the photosphere
should not necessarily be greater than the local escape velocity,
because periodic passage
of shock waves distends the stellar atmosphere and,
under some conditions, may lead to the gradual approach of the gas flow
to the escape velocity \citep{Willson:1979}.
Furthermore, the shock amplitude tends to increase while the shock wave
propagates in a medium with decaying gas density.

The most important problem concerning the shock--driven mass loss
is the rate of the shock decay during its passage through the stellar
atmosphere.
Various simplifying assumptions lead to estimates of mass loss rates
that differ from one another by many orders of magnitude \citep{Wood:1979}.
Therefore, an analysis must be based on the self--consistent solution
of the radiation transfer, fluid dynamics and rate equations.
Unfortunately, the description of the shock wave passage in terms of the
Lagrangean approach cannot be applied, since it does not allow the postshock
relaxation zones to be explicitly resolved with respect to the spatial
coordinate \citep{Klein:1976,Klein:1978}.
Another approach is based on the assumption of steady--state gas flow.
Applicability of this assumption for shock waves in stellar atmospheres
is justified by the small width
(including both the radiative precursor
and the postshock relaxation zone)
of the shock wave in comparison with
the atmosphere scale height.

Another necessary condition for applicability of the steady--state assumption
implies that the characteristic time scale of the shock wave decay is
$t_\mathrm{dec}\ll d/U$, where $d$ and $U$ are the width and velocity
of the shock wave.
In atmospheres of radially pulsating stars this condition is obviously
fulfilled because emission lines indicating
the shock wave are observed during the major part of the pulsation
period \citep{Abt:1954,Willson:1976} and in some stars this interval
is as long as a half \citep{Fox:1984} or even 0.8
\citep{Gillet:1985} of the pulsation period.

Even in the framework of the steady--state assumption the problem of the shock
wave structure in the partially ionized gas remains tremendously difficult,
due to the strong coupling between the radiative precursor and the shock wake.
Indeed, the occupation number densities of bound and free levels
in both atoms and molecules strongly depend on the radiation field produced
in the shock wake and at the same time must be determined as a solution
of rate equations that are stiff on the relevant time scales.
Fortunately, in stars with effective temperatures of $\Teff\gtrsim 3000\K$,
the gas is mostly in the atomic state and the principal dissipative processes
in the postshock region are excitation and ionization of atoms.

In Paper~I \citep{Fadeyev:1998} we proposed the method
of global iterations that allows us to obtain the stable self--consistent
solution of the equations of fluid dynamics, radiation transfer and
rate equations for the steady--state plane--parallel shock wave
propagating through the atomic hydrogen gas.
In Paper~II \citep{Fadeyev:2000} this method was employed for
computations of the structure of shocks with upstream velocities of
$15~\kms\le U_1\le 70~\kms$ and for the temperature and the density
of the unperturbed hydrogen gas of $T_1 = 6000\K$ and
$\rho_1 = 10^{-10}~\gcc$, respectively.
Results of these calculations are obviously insufficient for astrophysical
applications since the temperature and the density in atmospheres of pulsating
stars vary in wide ranges.

Below we discuss the structure of shock waves propagating through
partially ionized hydrogen gas with temperature and density ranges within
$3000\K\le T_1\le 8000\K$ and $10^{-12}~\gcc\le\rho_1\le 10^{-9}~\gcc$,
respectively.
Basic equations and the method of their solution are described in Paper~II.
In comparison with our previous work, we improved the treatment
of radiation transfer and extended the total spectral range
to $13\le\log\nu\le 16$.
Furthermore, we employed the much more correct treatment of the spectral line
radiation transfer which is of great importance for atomic level
populations obtained as solution of rate equations.
All calculations were done for the hydrogen atom with $L=4$ bound levels and
a continuum.

\section{General description of the shock wave structure}

In the comoving fluid frame the radiation--modified Rankine--Hugoniot
relations are written \citep{Marshak:1958,Mihalas:1984} as
\begin{equation}
\label{CC0}
\rho U = \CC_0\equiv \dot m ,
\end{equation}
\begin{equation}
\label{CC1}
\dot mU + \Pg + \PR = \CC_1 ,
\end{equation}
\begin{equation}
\label{CC2}
\frac{1}{2}\dot m U^2 + \dot m h + \FR + U\left(\ER + \PR\right) = \CC_2 ,
\end{equation}
where
$\rho$ is the gas density, $U$ is the gas flow velocity,
$h$ is the specific enthalpy, $\Pg$ is the gas pressure,
$\ER$, $\FR$ and $\PR$ are the radiation energy density, radiation flux
and radiation pressure, respectively.

In planar geometry the total energy flux $\CC_2$ given by relation (\ref{CC2})
is constant along the spatial coordinate $X$, so that while the parcel
of gas passes the shock wave the decrease of the
flux of kinetic energy $\FK = \frac{1}{2}\dot mU^2$
is balanced by changes in the enthalpy flux
$\Fh = \dot mh$ and the radiation flux $\FR$.
Throughout the shock wave the flux $U(\ER+\PR)$ is significantly
smaller than other terms of the left--hand side of
relation (\ref{CC2}) and can be ignored.

As in Paper~II the shock wave model is represented by the comoving
plane--parallel finite slab.
The space coordinate $X=0$ is set at the viscous adiabatic jump which
is treated as an infinitesimally thin discontinuous jump where
hydrodynamic variables undergo an abrupt change.
It is assumed that the space coordinate $X$ of the gas element increases
while the gas flows through the shock wave, so that in the preshock region
$X<0$ and in the postshock region $X>0$.
The coordinates of the preshock and postshock outer boundaries are denoted as
$X_1$ and $X_N$, respectively.
The upstream radiation flux emerging from the preshock outer boundary
is negative, whereas the downstream radiation flux emerging from the postshock
outer boundary is positive, that is, $\FRa < 0$ and $\FRn > 0$.

\begin{figure}
\resizebox{\hsize}{!}{\includegraphics{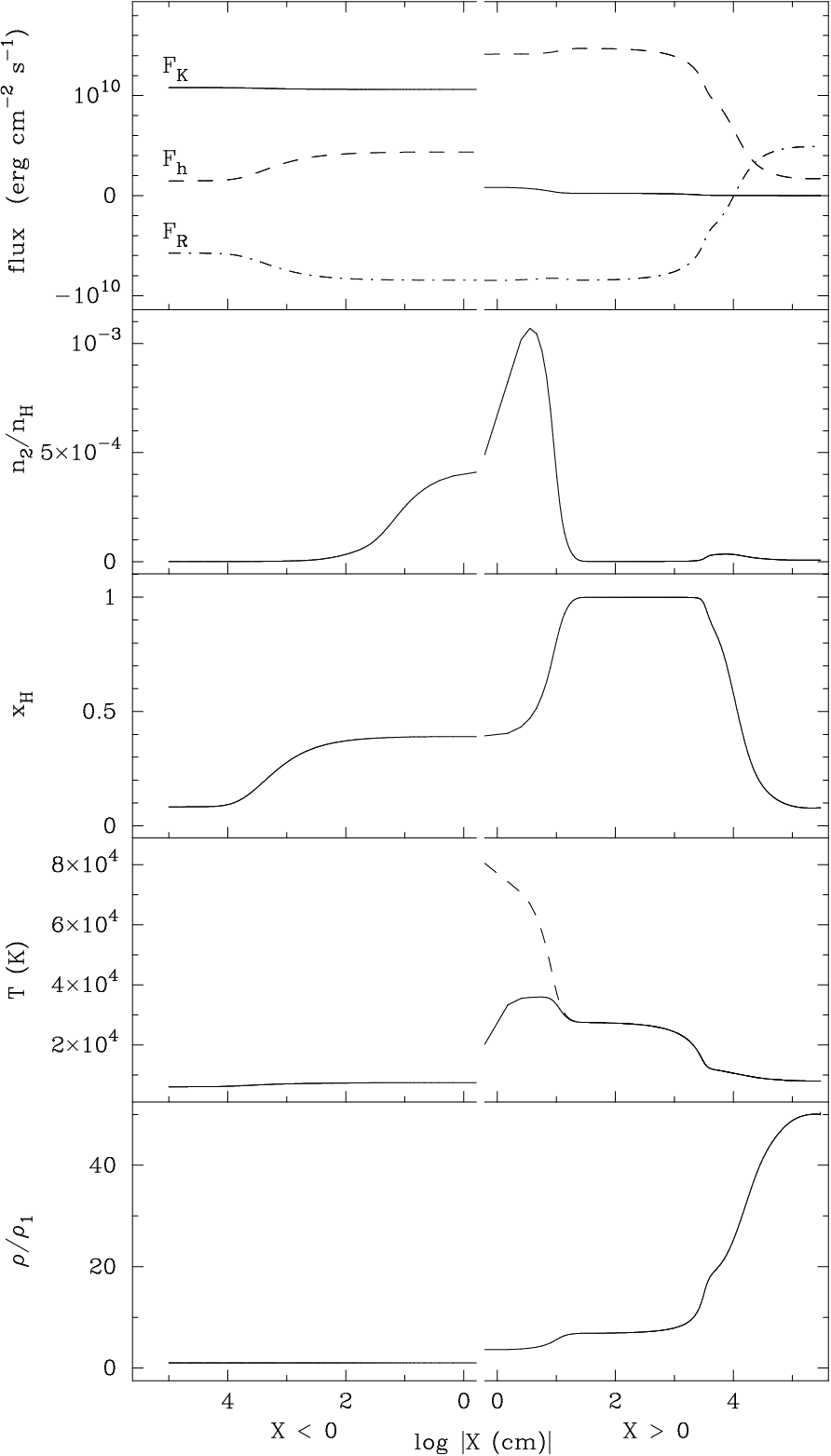}}
\caption{The structure of the shock wave with
$\rho_1=10^{-10}~\gcc$, $T_1=6000\K$ and $U_1=60~\kms$.
On the upper panel the solid lines, dashed lines and dot--dashed lines
represent the flux of the kinetic energy $\FK$, the flux of the enthalpy $\Fh$
and the radiation flux $\FR$.
On lower panels are given the fractional number density of hydrogen atoms
in the 2--nd state $n_2/\nH$,
the hydrogen ionization degree $\xH$,
the electron temperature $\Te$ (solid lines) and the temperature of heavy
particles $\Ta$ (dashed line),
and the compression ratio $\rho/\rho_1$.}
\label{flux}
\end{figure}

On the upper panel of Fig.~\ref{flux}, we show the fluxes
$\FK$, $\Fh$ and $\FR$ as a function of space coordinate $X$
for the shock wave model with 
$\rho_1=10^{-10}~\gcc$, $T_1=6000\K$ and $U_1=60~\kms$.
For the sake of convenience we use the logarithmic scale along the spatial
coordinate $X$, the preshock and postshock regions being represented
by the left plot and the right plot, respectively.
Below the upper panel of Fig.~\ref{flux}, we show as a function of $X$
the fractional number density of hydrogen atoms in 2--nd state $n_2/\nH$,
the hydrogen ionization degree $\xH=\nel/\nH$,
the electron temperature $\Te$ and the temperature of heavy particles
(i.e. of neutral hydrogen atoms and hydrogen ions) $\Ta$,
and on the lowest panel we give the plot of the compression ratio $\rho/\rho_1$.

We assume that at the preshock outer boundary $X_1$, the gas temperature
and the gas density are the same as those of the unpertubed medium
that are denoted as $T_1$ and $\rho_1$, respectively.
At the same time, the upstream radiation flux emerging from the shock wake
affects both the bound--level number densities and the ionization degree
of hydrogen atoms.
Therefore, the number densities of hydrogen atoms in $i$--th state $n_i$
and the number density of free electrons $\nel$ at the preshock outer
boundary are determined from the solution of the equations of statistical
equilibrium.
In our calculations we tried to set the preshock outer boundary $X_1$ as far
as possible from the discontinuous jump in order to diminish the perturbing
influence of the shock wave radiation field on the preshock gas.

The most prominent changes that the gas flow undergoes in the preshock region
are due to the absorption of the Lyman continuum radiation.
In Fig.~\ref{flux} the radiative precursor is revealed by the growth
of the enthalpy flux $\Fh$ and due to the corresponding changes
of the radiation flux $\FR$.
The length of the radiative precursor $\delta X_\mathrm{p}$ is
proportional to
the mean free path of Lyman continuum photons and increases with decreasing
density from $\delta X_\mathrm{p}\sim 10^3~\cm$ at $\rho_1=10^{-9}~\gcc$
to $\delta X_\mathrm{p}\sim 10^7~\cm$ at $\rho_1=10^{-12}~\gcc$.

Within the radiative precursor the divergence of the Lyman continuum flux is
negative, that is,
\begin{equation}
\nabla\!\cdot\mathbf{F}_\mathrm{Lyc} =
4\pi\int\limits_{\nu_1}^\infty
\left(\eta_\nu - \kappa_\nu J_\nu\right) d\nu < 0 ,
\end{equation}
where $\eta_\nu$ and $\kappa_\nu$ are the monochromatic emission and absorption
coefficients, $J_\nu$ is the mean intensity and $\nu_1$ is the threshold
frequency for ionization from the ground state of the hydrogen atom.
Thus, the gas temperature increases due to absorption of the Lyman continuum
radiation while the gas parcel approaches the discontinuous jump.
As a result, the gas density $\rho$ just ahead of the discontinuity
exceeds its unperturbed value $\rho_1$, whereas the gas
flow velocity $U$ becomes smaller than $U_1$.
The change of the gas density $\rho$ and the gas flow velocity $U$ in the
radiative precursor depends not only on the upstream velocity $U_1$
but also on the temperature and density of the unperturbed preshock gas.
At $\rho_1=10^{-10}~\gcc$ and $U_1=75~\kms$, the relative growth of the
density within the radiative precursor ranges from
$\Delta\rho/\rho_1\approx 0.03$ at $T_1=3000\K$ to
$\Delta\rho/\rho_1\approx 0.07$ at $T_1=7000\K$.

It should be noted that when the hydrogen gas is not completely
ionized in the preshock region, the increase in the enthalpy flux is mostly
due to ionization of hydrogen atoms.
The rates of collisional transitions, both bound--bound and bound--free,
are negligible in comparison with those of radiative transitions, so that
the increase in $\Fh$ ahead of the discontinuous jump results mostly from the
photoionization of hydrogen atoms.

Across the discontinuity, the flux of the kinetic energy $\FK$ undergoes
an abrupt decrease which is exactly balanced by the same abrupt increase
of the enthalpy flux $\Fh$.
Because population number densities per unit mass
$n_i/\rho$ and $\nel/\rho$ do not change across the discontinuity, 
the increase in the specific enthalpy is only due to the
abrupt increase in the temperature of heavy particles $\Ta$ and
the electron temperature $\Te$.
Here we assume that the temperature of heavy particles just behind the
discontinuous jump is given by the solution of the Rankine--Hugoniot relations
(\ref{CC0}) -- (\ref{CC2}):
\begin{eqnarray}
\label{Tab}
\Ta^+ &=& \Ta^- - \xH^- \left(\Te^+ - \Te^-\right) +
\frac{1}{5}\frac{\dot m U^-}{\nH^- k}
\frac{\eta^2 - 1}{\eta^2} -
\nonumber \\
&-&\frac{2}{5}\frac{\FR^+ - \FR^-}{\nH^- kU^-} -
\nonumber \\
&-&\frac{2}{5}
\frac{(\ER^+ + \PR^+)\eta^{-1} - (\ER^- + \PR^-)}{\nH^- k} ,
\end{eqnarray}
whereas the electron temperature increases due to adiabatic compression:
\begin{equation}
\Te^+ = \eta^{\gamma - 1}\Te^- ,
\end{equation}
where $\eta=\rho^+/\rho^-$ is the compression ratio across the discontinuous
jump, $\gamma=5/3$ is the ratio of specific heats of the electron gas,
and superscripts minus and plus refer to the variables defined ahead of and 
behind the discontinuous jump, respectively
\footnote{In relation (\ref{Tab}) we corrected the typeset errors which 
appeared in relation (15) of Paper~II.}.

The mean free path of photons in the vicinity
of the discontinuous jump is several orders of magnitude larger
than that of hydrogen atoms and the radiation flux will remain constant
to a high order of accuracy, even if we take into account the finite thickness
of the viscous adiabatic jump.
In our calculations, the space interval between two adjacent cell centers
located ahead of and behind the discontinuity was set equal to
$1~\cm$ for all models. 
Thus, the last two terms on the r.h.s. of
(\ref{Tab}) can be omitted
without loss of accuracy because they refer to the difference between the
upstream and downstream values of the radiation energy density, radiation flux
and radiation pressure.

Excitation and ionization of hydrogen atoms behind the discontinuity are
mostly due to radiative transitions.
The timescale of photoexcitation of the lower bound levels
is comparable with that of the electron temperature growth, whereas
the photoionization of hydrogen atoms is much slower and begins with
a substantial delay.
The increase in the hydrogen ionization leads to a small increase in the
enthalpy flux and, correspondingly, to the small decrease in the
flux of kinetic energy because the radiation flux remains almost constant.
However, the most promiment process accompanying hydrogen ionization is
the increase in the compression ratio $\rho/\rho_1$.
For the model presented in Fig.~\ref{flux}, the compression ratio
increases from $\rho/\rho_1=3.6$ just behind the discontinuity, to
$\rho/\rho_1=6.8$ in the layers where the hydrogen ionization reaches
its maximum.

At larger distances from the discontinuity the kinetic energy of the gas flow
initially stored as the ionization energy of the gas is converted into 
radiation.
In Fig.~\ref{flux} these layers are revealed as those of remarkable changes
in $\Fh$ and $\FR$, accompanied by the growth of the
compression ratio $\rho/\rho_1$.
For models considered in the present study, the compression ratio
reached at the postshock outer boundary ranges from
$1 < \rho_N/\rho_1\lesssim 10^2$ depending mostly on the upstream
velocity $U_1$ and less on the preshock temperature $T_1$ and preshock
density $\rho_1$.

The large increase in the postshock compression ratio
$\eta = \rho/\rho_1$ is due to the ionization of hydrogen gas and its
subsequent radiative cooling.
In order to roughly estimate the contribution of these effects we can write the
energy conservation relation (\ref{CC2}) with the omitted radiation energy 
density and radiation pressure terms as
\begin{equation}
\label{CC2a}
\begin{array}{l}
\dsfrac{U_1^2}{2} + \Etra + \Eina + \dsfrac{\Pga}{\rho_1} +
\dsfrac{\FRa}{\dot m} =
\\
= \dsfrac{U^2}{2} + \Etr + \Ein + \dsfrac{\Pg}{\rho} + \dsfrac{\FR}{\dot m} ,
\end{array}
\end{equation}
where $\Etr = \frac{3}{2}(\nH+\nel)kT/\rho$ is the specific energy in
the translational degrees of freedom and $\Ein$ is the specific
energy of excitation and ionization of hydrogen atoms.
The left--hand side of (\ref{CC2a}) is written for the preshock outer
boundary, whereas its right--hand side represents an arbitrary layer of
the postshock region.

Relation (\ref{CC2a}) and the momentum conservation relation (\ref{CC1})
can be rewritten as
\begin{equation}
\label{eta}
\eta = 4 + 3\frac{\Ein - \Eina}{\Etr} + 3\frac{\FR - \FRa}{\dot m\Etr} ,
\end{equation}
where we assumed that the specific energy
in the translational degrees of freedom of the preshock gas
is negligible in comparison with that of the postshock compressed gas,
that is, $\Etra/\eta\ll\Etr$.

The first term in the right--hand side of (\ref{eta}) is the limiting
compression ratio of the atomic hydrogen gas at the discontinuous jump.
The second term in the right--hand side of (\ref{eta}) describes the
compression of the postshock gas due to excitation and ionization of
hydrogen atoms.
For the model shown in Fig.~\ref{flux}, the maximum hydrogen ionization
degree is $\xH=0.99$ and the ratio of the specific energies is
$\Ein/\Etr\approx 2$.
Thus, the upper limit for the compression ratio beyond the layers of hydrogen
ionization is $\eta\approx 10$.

However the most important contribution to the growth of the compression
ratio comes from the third term in the right--hand side of (\ref{CC2a}).
Indeed, for the model shown in Fig.~\ref{flux} the ratio of the sum
of radiative fluxes (note that $\FRa$ is negative) to the flux of the
translational energy of the gas at the postshock outer boundary is
$(\FRn - \FRa)/(\dot m\Emn)\approx 18$.
Therefore, the maximum compression ratio at the postshock outer boundary is
$\eta\approx 64$.

\section{Effects of the preshock temperature}

Just behind the discontinuity, the temperature of electrons, $\Te$, is
substantially lower than the temperature of heavy particles, $\Ta$, and
electrons acquire energy in elastic collisions with neutral hydrogen
atoms and hydrogen ions.
Because the elastic scattering crosssection of electrons and hydrogen ions
is much larger than that of electrons and neutral hydrogen atoms,
the efficiency of the energy exchange between
heavy particles and free electrons is very sensitive to the hydrogen
ionization degree, $\xH^-$, ahead of the discontinuous jump.
In particular, at $\xH^-\gtrsim 10^{-2}$, the postshock equilibration between 
the electron temperature $\Te$ and the temperature of heavy particles $\Ta$
is due to elastic scattering of electrons by hydrogen ions and the equilibration
rate very rapidly increases with increasing $\xH^-$.
Also, in reverse, at $\xH < 10^{-2}$, free electrons gain most of the
energy from elastic collisions with neutral hydrogen atoms and
the rate of temperature equilibation is much slower.
This process, however, is perceptible only in shock waves with weak
ionization in the
radiative precursor (that is, at upstream velocities $U_1 < 30~\kms$),
the preshock gas temperature being $T_1 < 4000\K$.

\begin{figure}
\resizebox{\hsize}{!}{\includegraphics{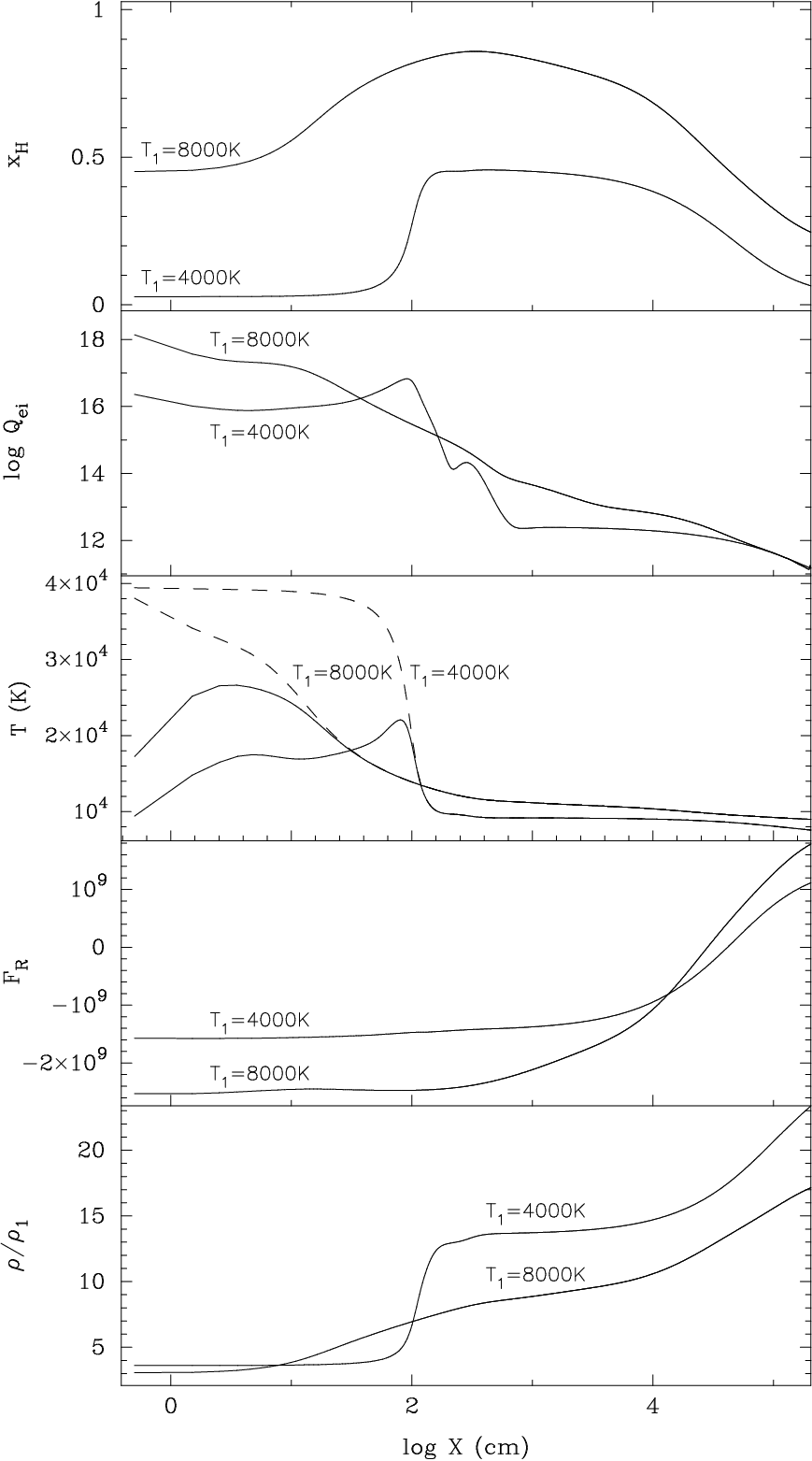}}
\caption{The postshock hydrogen ionization degree $\xH$,
the rate of energy gain by electrons in elastic collisions with
hydrogen ions per unit mass $Q_\mathrm{ei}$,
the electron temperature $\Te$ (solid lines) and
the temperature of heavy particles $\Ta$ (dashed lines);
the radiation flux $\FR$
and the compression ratio $\rho/\rho_1$ as a function of distance from
the discontinuous jump in shock wave models with $\rho_1=10^{-10}~\gcc$,
$U_1=40~\kms$, $T_1=4000\K$ and $T_1=8000\K$.}
\label{r100u40}
\end{figure}

Thus, effects of the preshock gas temperature on the shock wave structure are
mostly due to the preshock equilibrium hydrogen ionization.
This is illustrated in Fig.~\ref{r100u40} for shock wave models with
$T_1=4000\K$ and $T_1=8000\K$.
In both cases, the preshock gas density and the upstream velocity are
$\rho_1=10^{-10}~\gcc$ and $U_1=40~\kms$, respectively, so that the postshock
equilibration
of $\Te$ and $\Ta$ is due to elastic scattering of electrons by hydrogen ions.
As is seen from the upper panel of Fig.~\ref{r100u40}, the hydrogen ionization
degree just behind the discontinuous jump is $\xH\approx 0.02$ at
$T_1=4000\K$ and $\xH\approx 0.46$ at $T_1=8000\K$.
As a result, at a preshock gas temperature $T_1=8000\K$,
the rate of the energy gain by electrons in elastic collisions
with hydrogen ions is larger by a factor of 60 than that
at $T_1=4000\K$.

The increase in the preshock gas temperature $T_1$ is accompanied by
the stronger radiation flux emerging from the shock wave.
This is, obviously, due to the higher hydrogen ionization
degree in the postshock region.
The more gradual changes in the postshock electron temperature and of the
ionization degree in shocks with higher preshock temperature
lead to smaller compression ratios $\rho/\rho_1$
(see the lower panel of Fig.~\ref{r100u40}).

\begin{figure}
\resizebox{\hsize}{!}{\includegraphics{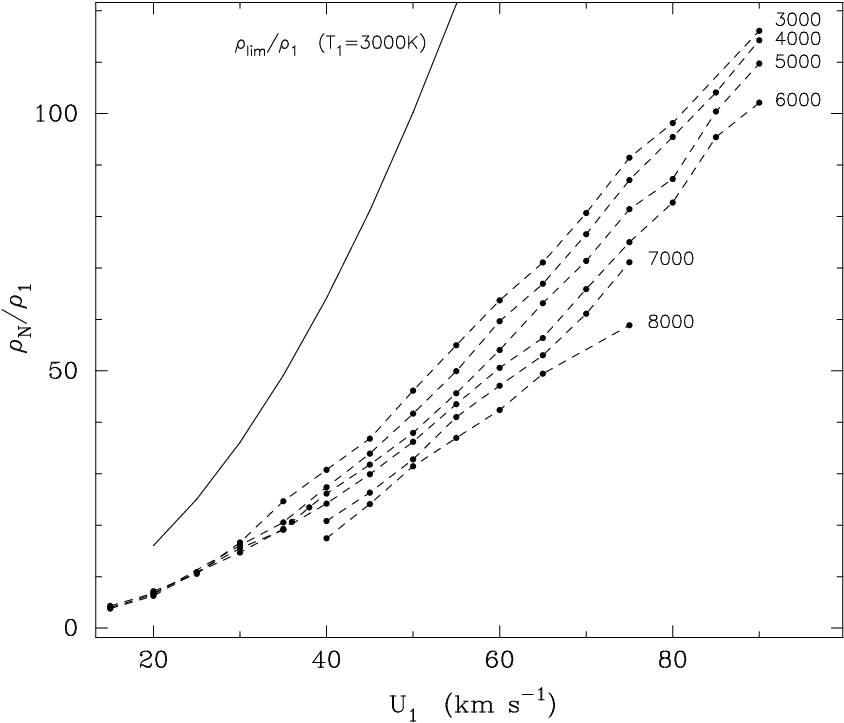}}
\caption{The compression ratio $\rho_N/\rho_1$ at the postshock outer
boundary as a function of upstream velocity $U_1$
at $\rho_1=10^{-10}~\gcc$ and $3000\K\le T_1\le 7000\K$.
Each sequence of models with the same preshock temperature $T_1$ is represented
by filled circles connected by dashed lines.
Each curve is labeled with $T_1$.
The solid line shows the compression ratio of the isothermal shock wave
propagating through the gas with $T_1=3000\K$.}
\label{rhon}
\end{figure}

In Fig.~\ref{rhon} we give the plots of the final compression ratio
$\rho_N/\rho_1$ which is reached at the postshock outer boundary $X_N$
in shock wave models with $\rho_1=10^{-10}~\gcc$.
It should be noted that the final postshock gas density asymptotically tends
to its limiting value $\rho_\infty$ with $X\to\infty$ and
because in our study the shock wave model is represented by the finite slab,
the compression ratio $\rho_N/\rho_1$ gives only the lower estimate
for the limiting value $\rho_\infty/\rho_1$.
In our calculations we tried to set the postshock outer boundary as far as
possible from the discontinuous jump.
Unfortunately, the convergence of global iterations
is very sensitive to the spatial coordinate of the postshock outer boundary
and for too large values of $X_N$, the iterations diverge.
Thus, the coordinate $X_N$ was determined for each model from trial
calculations as a compromise between the requirement of
the postshock region and demands of convergence and accuracy.
For this reason, the filled circles in Fig.~\ref{rhon}, representing the models
with the same value of $T_1$, do not lie on smooth curves.

\begin{figure}
\resizebox{\hsize}{!}{\includegraphics{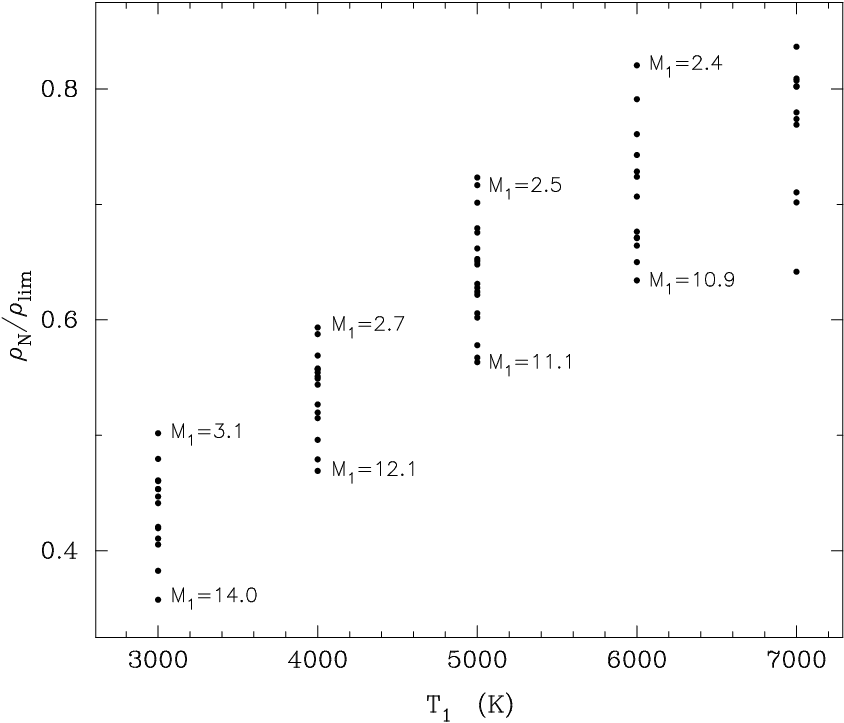}}
\caption{The final compression ratio $\rho_N/\rho_1$ at the postshock outer
boundary in units of the compression ratio of the isothermal shock wave
$\etalim$ versus the preshock gas temperature $T_1$ for models with
$\rho_1=10^{-10}~\gcc$ and $20~\kms\le U_1\le 90~\kms$.
At fixed preshock temperature $T_1$ the values of the Mach number $M_1$
correspond to models with $U_1=20~\kms$ and $U_1=90~\kms$, respectively.}
\label{rholim}
\end{figure}

At a fixed upstream velocity, $U_1$, the Mach number $M_1=U_1/a_1$ decreases 
with increasing $T_1$ due to the temperature dependence of the
adiabatic sound speed $a_1$ at the preshock outer boundary.
In a hydrogen gas with negligible ionization
($\rho_1=10^{-10}~\gcc$, $T_1\lesssim 7000\K$)
the adiabatic sound speed is $a_1\propto\sqrt{T_1}$ and the compression
ratio
\begin{equation}
\etalim = \gamma M_1^2
\end{equation}
corresponding to the isothermal shock wave is inversly proportional to $T_1$.
On the other hand, as is seen from Fig.~\ref{rhon},
the final compression ratio $\rho_N/\rho_1$ at the postshock outer boundary
also decreases with increasing preshock temperature though this dependence
is not as prominent as that of $\etalim$.

In Fig.~\ref{rholim} we show the ratio $\rho_N/\rholim$ for shock wave
models with $\rho_1=10^{-10}~\gcc$,
$3000\K\le T_1\le 7000\K$, $20~\kms\le U_1\le 90~\kms$.
Each model on this plot is represented by a filled circle and for
models with $U_1=20~\kms$ and $U_1=90~\kms$ we give the values
of the upstream Mach number $M_1$.

At a fixed preshock gas temperature $T_1$ the ratio
$\rho_N/\rholim$ decreases with increasing upstream velocity since
the final compression ratio $\rho_N/\rho_1$ grows with $U_1$
more slowly than $\etalim$ (compare dependencies of $\rho_N/\rho_1$ and
$\rholim/\rho_1$ shown for $T_1=3000\K$ in Fig.~\ref{rhon}).
However the most interesting feature is that the increase in the preshock
gas temperature $T_1$ is accompanied by the gradual approach of the
final compression ratio $\rho_N/\rho_1$ to its upper limit
$\etalim$ corresponding to the isothermal shock wave.
This is due to the decrease of the Mach number $M_1$
with increasing $T_1$.

\section{Effects of the preshock density}

\begin{figure}
\resizebox{\hsize}{!}{\includegraphics{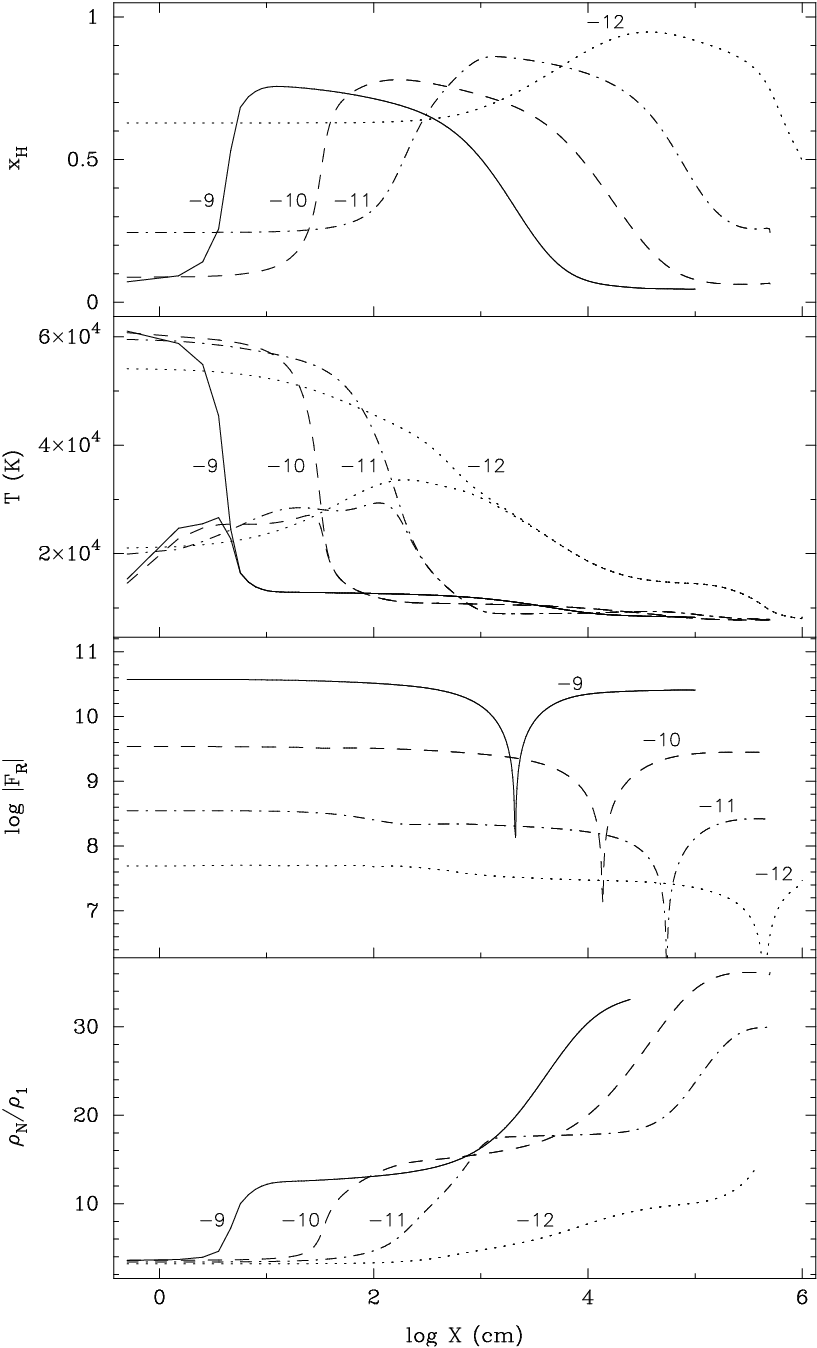}}
\caption{The hydrogen ionization degree $\xH$,
the electron temperature $\Te$ and the temperature of heavy particles $\Ta$,
the radiation flux $\FR$ and the compression ratio $\rho/\rho_1$
as a function of distance from the discontinuous jump
in the postshock region at $T_1=6000\K$ and $U_1=50~\kms$.
In solid, dashed, dot--dashed and dotted lines are shown
the dependencies corresponding to the preshock gas densities
$\rho_1=10^{-9}$, $10^{-10}$, $10^{-11}$ and $10^{-12}~\gcc$, respectively.
Each curve is labeled with $\log\rho_1$.
In temperature plots, the lower and upper converging curves represent
the electron temperature $\Te$ and the heavy particle temperature $\Ta$,
respectively.
}
\label{t6u50}
\end{figure}

The two upper panels of Fig.~\ref{t6u50} show the postshock hydrogen ionization
degree $\xH$ and the postshock temperatures $\Te$ and $\Ta$
as a function of distance from the discontinuous jump for shock wave
models with $10^{-12}~\gcc\le\rho_1\le10^{-9}~\gcc$,
$T_1=6000\K$ and $U_1=50~\kms$.
The two lower panels of Fig.~\ref{t6u50} for the same models show
the radiation flux $\FR$ and the compression ratio $\rho/\rho_1$.
Note that we use the logarithmic scale for the plot of radiative flux and
the deep minima of $\log|\FR|$ correspond to layers with $\FR\approx 0$.

The main results of these calculations are as follows.
First, the width of the postshock relaxation zone increases with
decreasing preshock gas density $\rho_1$.
This is due to the fact that the mean collision time of particles is
proportional to the gas density.
Second, the radiative flux produced by the shock wave decreases with
decreasing $\rho_1$ because the kinetic energy of the gas flow
at the preshock outer boundary is proportional to $\rho_1$.
Third, more gradual ionization and recombination of hydrogen atoms
at the lower preshock gas density $\rho_1$ leads to smaller postshock
compression ratios.

\section{Radiative flux of the shock wave}

As seen from plots on the upper panel of Fig.~\ref{flux},
the upstream radiative flux $\FRa$ emerging from the preshock outer boundary
and the downstream radiation flux $\FRn$ emerging in the opposite direction 
from the postshock outer boundary are not equal in absolute value.
This difference increases with increasing upstream velocity.
For example, at $\rho_1=10^{-10}~\gcc$ and $T_1=6000\K$
the ratio of these fluxes ranges
from $|\FRa|/\FRn \approx 1.01$ at $U_1=20~\kms$ to $|\FRa|/\FRn \approx 1.16$
at $U_1 = 60~\kms$.

However the much stronger asymmetry of the radiation field is revealed at
frequencies $\nu > \nu_1$ because the Lyman continuum flux is directed
mostly upstream.
The region of the effective transport of the Lyman continuum radiation
encompasses the near vicinity of the discontinuous jump.
Ahead of the discontinuity, this is the radiative precursor and behind
the discontinuity the Lyman continuum flux becomes negligible
where the total radiation flux changes its sign (i. e.: in the layers
with $\FR\approx 0$).
The contribution of the Lyman continuum to the total radiation flux
within this region rapidly increases with increasing upstream velocity.
For example, for shock waves propagating in a gas with
$\rho_1=10^{-10}~\gcc$ and $T_1=3000\K$,
the ratio of the Lyman continuum flux to the total radiation flux
increases from $\FLyc/\FR\approx 10^{-2}$ at $U_1=40~\kms$
to $\FLyc/\FR\approx 0.2$ at $U_1=60~\kms$.
The growth of the ratio $\FLyc/\FR$ ceases at an upstream velocity
$U_1 \approx 75~\kms$, corresponding to almost full hydrogen ionization
in the radiative precursor.
At larger upstream velocities the fraction of the Lyman continuum radiation
does not exceed $\approx 40\%$ of the total radiation flux, but the
region of the effective transport of the Lyman continuum radiation becomes
wider, both ahead of and behind the discontinuity.

\begin{figure}
\resizebox{\hsize}{!}{\includegraphics{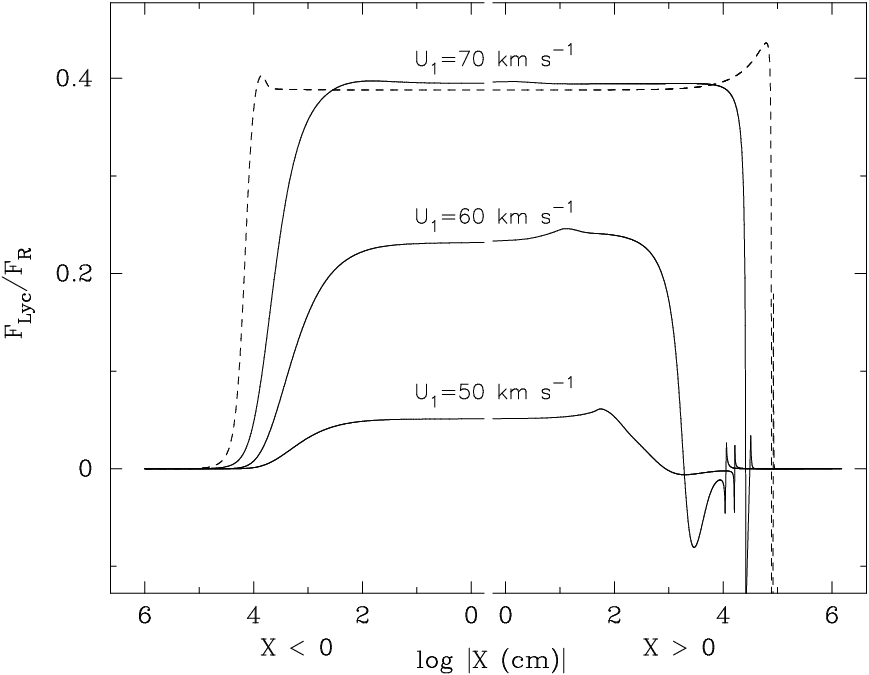}}
\caption{The ratio of the radiation flux in the Lyman continuum
to the total radiation flux $\FLyc/\FR$ in the vicinity of the discontinuous
jump of shock waves with an upstream velocity $U_1=50$, 60, 70
(solid lines) and $80~\kms$ (dashed line).}
\label{frati1}
\end{figure}

The plots of the ratio $\FLyc/\FR$ for shock wave models with upstream
velocities $50~\kms\le U_1\le80~\kms$ are shown in Fig.~\ref{frati1}.
It should be noted that the positive value of the ratio $\FLyc/\FR$
implies that the total radiation flux and the Lyman continuum
flux are both negative and are directed upstream.
Negative values of this ratio imply that the Lyman continuum flux is
directed downstream, whereas the total radiation flux is still upstream.
Oscillations of the ratio $\FLyc/\FR$ in the postshock region are due to 
vanishing values of $\FR$.

\begin{figure}
\resizebox{\hsize}{!}{\includegraphics{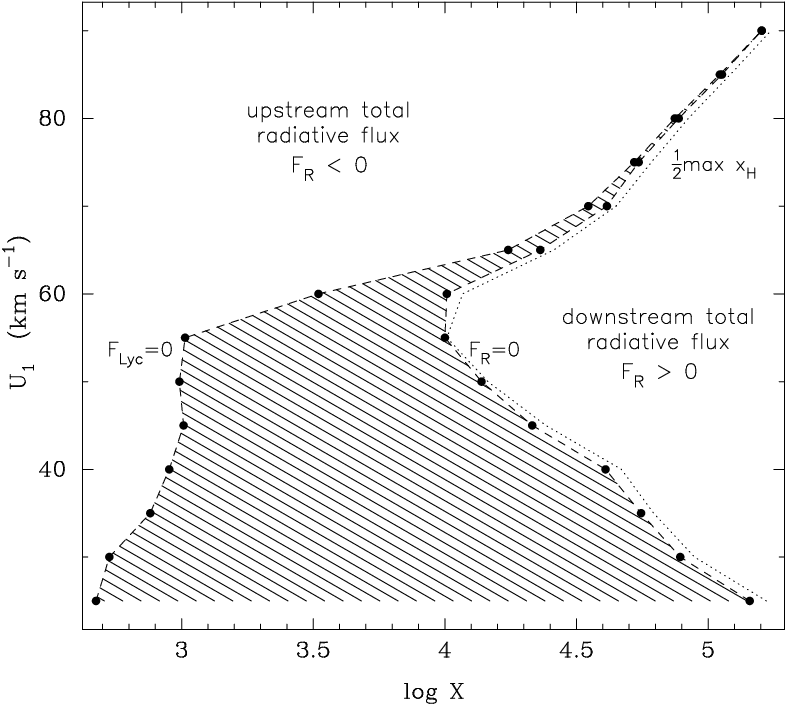}}
\caption{The zones of upstream ($\FR < 0$) and downstream ($\FR > 0$)
total radiative flux in the postshock region at
$\rho_1=10^{-10}~\gcc$ and $T_1=6000\K$.
The shaded area represents the zone with downstream Lyman
continuum flux and upstream flux $F(\nu < \nu_1)$.
The dotted curve marked $\frac{1}{2}\max\xH$ gives the locus where
the hydrogen ionization degree is a half of its maximum postshock value.}
\label{xfr0}
\end{figure}

Thus, behind the discontinuous jump, the Lyman continuum flux $\FLyc$
changes its sign at smaller distance than the radiative flux
\begin{equation}
F(\nu < \nu_1) = \int\limits_0^{\nu_1} F_\nu d\nu\,,
\end{equation}
transported at frequencies below the Lyman continuum edge $\nu_1$.
In Fig.~\ref{xfr0} we depict the zones of
the postshock region with upstream and downstream total radiative flux
$\FR$ in shock wave models with $\rho_1=10^{-10}~\gcc$ and $T_1=6000\K$.
The zone with downstream Lyman continuum flux ($\FLyc > 0$) and
upstream flux $F(\nu < \nu_1)$ is shown on this diagram by the
shaded area.
Space coordinates $X[F(\nu < \nu_1) = 0]$ and $X(\FR=0)$ coincide
because most of the radiation is transported in these layers at
frequencies below the Lyman continuum edge.

In Fig.~\ref{xfr0} we also show the space coordinates of the layers where
half of the ionized hydrogen atoms are recombined.
The close values of $X(\FR=0)$ and $X(\frac{1}{2}\max\xH)$ clearly
illustrate that the main mechanism of the shock wave energy dissipation
is the ionization of hydrogen atoms.

At upstream velocity $U_1 < 60~\kms$, the postshock gas remains partially
ionized and the maximum ionization degree increases with increasing $U_1$,
while the space coordinate of the ionization maximum decreases.
Correspondingly, the hydrogen recombination also occurs at a smaller distance
from the discontinuous jump.

At upstream velocities of $U_1 > 60~\kms$, the hydrogen is fully ionized
and the width of the zone with $\xH\approx 1$ increases with increasing $U_1$.
Correspondingly, the hydrogen recombines at larger distances from the
discontinuous jump and the coordinate $X(\FR=0)$ increases with increasing
$U_1$. At full hydrogen ionization, the zone of the  effective energy
transport in the Lyman continuum spreads both upstream and downstream
with increasing upstream velocity and at $U_1 > 80~\kms$ both the Lyman
continuum flux and the total radiation flux change sign nearly
in the same layers.

Another interesting conclusion which follows from the diagram in Fig.~\ref{xfr0}
is that the width of shock waves with partial postshock hydrogen ionization
decreases with increasing upstream velocity $U_1$, whereas
the width of shock waves with full postshock hydrogen ionization increases with
increasing upstream velocity.

\section{Frequency dependent radiation field}

\begin{figure}
\resizebox{\hsize}{!}{\includegraphics{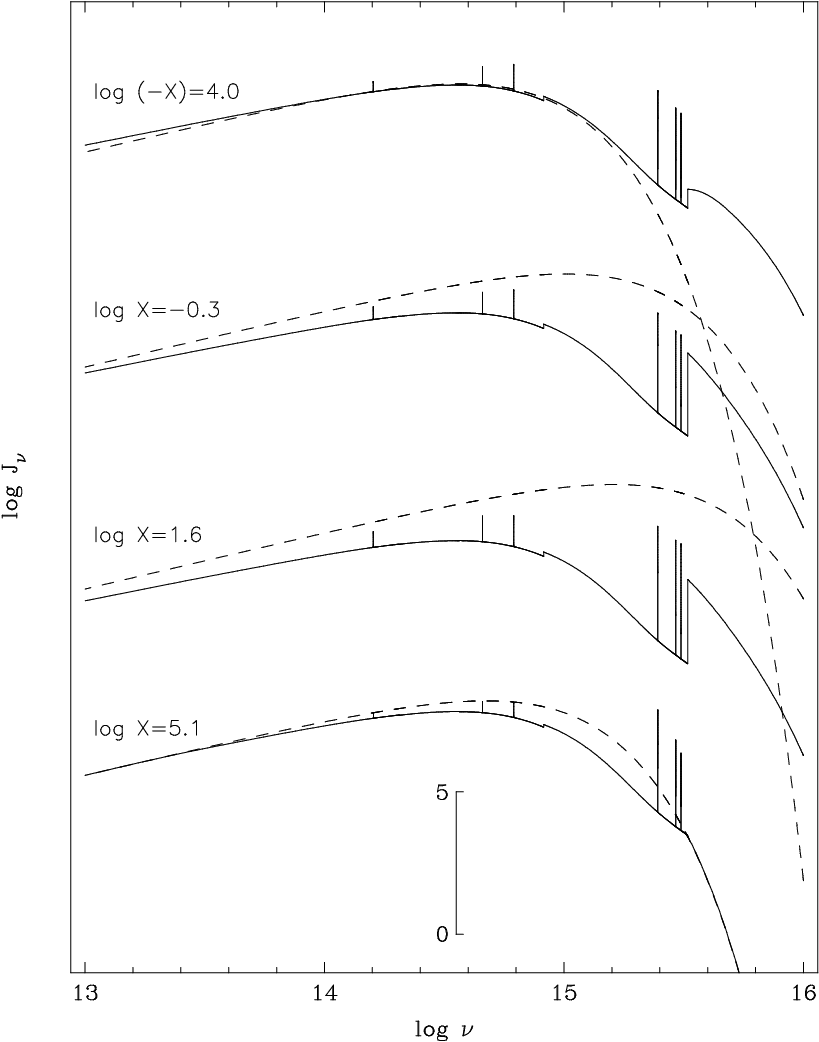}}
\caption{Monochromatic mean intensity $J_\nu$ (solid lines) and
the local Planck function $B_\nu(\Te)$ (dashed lines)
as a function of frequency $\nu$ for four distinct layers of the
shock wave with $\rho=10^{-10}~\gcc$, $T_1=6000\K$ and $U_1=60~\kms$.
The upper pair of plots represent the preshock region, whereas other
plots represent the postshock region.}
\label{jnu}
\end{figure}

The radiation field produced by the shock wave is remarkably
non--equilibrium and any attempts to use an assumption of thermal
equilibrium inevitably lead to large errors.
In order to evaluate the degree of departure from thermal
equilibrium, it is instructive to compare the monochromatic mean
intensity $J_\nu$ with the local Planck function $B_\nu(\Te)$.
In Fig.~\ref{jnu}  the plots of $J_\nu$ and $B_\nu(\Te)$ are shown as
a function of frequency $\nu$ for four distinct layers of the shock wave model
with $\rho=10^{-10}~\gcc$, $T_1=6000\K$ and $U_1=60~\kms$.
For the sake of convenience each pair of plots is shifted with respect
to others.

The upper pair of plots represents $J_\nu$ and $B_\nu(\Te)$ at
$X\approx 10^4~\cm$.
As is seen from Fig.~\ref{flux}, the photoionization of
hydrogen atoms in these layers due to absorption of the Lyman continuum
radiation becomes perceptible and this layer can roughly be considered as
the boundary of the radiative precursor.
Excess radiation in comparison with $B_\nu(\Te)$ at frequencies $\nu > \nu_1$
is due to the fact that the optical depth of the radiative precursor decreases
with increasing frequency $\nu$.

The second and third pairs of plots represent the cell just behind the
discontinuous jump ($X=0.5~\cm$) and the layer where the hydrogen ionization
degree reaches its maximum ($X\approx 40~\cm$).
Between these layers, various relaxation processes redistribute the
energy of heavy particles among other degrees of freedom and the radiation
field is furthest from equilibrium.

The lower pair of plots represent the layers ($X\approx 1.2\cdot 10^5~\cm$)
where most of the hydrogen atoms are recombined and the gas temperature
gradually approaches its unperturbed value $T_1$.
The optical depth at frequencies of the Lyman continuum is so large
that $J_\nu$ and $B_\nu(\Te)$ coincide at $\nu > \nu_1$.

All shock wave models demonstrate two common features.
First, we see as  excess of Lyman continuum radiation in the preshock
region, due to the existence of the radiative precursor.
Second, there is a lower mean intensity of $J_\nu$ in comparison with 
$B_\nu(\Te)$, due to the small optical depth of the shock wave at frequencies
$\nu < \nu_1$.

The plots of the mean intensity $J_\nu$ displayed in Fig.~\ref{jnu} reveal
the presence of six spectral line features.
These are \Lya, \Lyb, \Lyg, \Ha, \Hb\ and \Pa.
In comparison with our previous work described in Paper~II, we
improved the treatment of the spectral line radiation transfer
and considered each line of frequency $\nu_0$ within the frequency
interval as wide as
$[\nu_0(1-\delta),\nu_0(1+\delta)]$, where $\delta=5\cdot 10^{-4}$.
The use of wide frequency intervals is necessary because of the strong
Doppler broadening of line profiles behind the discontinuous jump.
In particular, the radiation flux transported
in spectral lines is not negligible and contribution of the spectral line
radiation to the total radiation flux increases with increasing upstream
velocity. The most important are the \Lya\ and \Ha\ lines.
For example, in the shock wave model with
$\rho_1=10^{-10}~\gcc$, $T_1=6000\K$ and $U_1=60~\kms$
the upstream radiation flux in \Lya\ and \Ha\ lines
emerging from the preshock outer boundary is $\approx 2\%$ of the total
radiation flux.

\begin{figure}
\resizebox{\hsize}{!}{\includegraphics{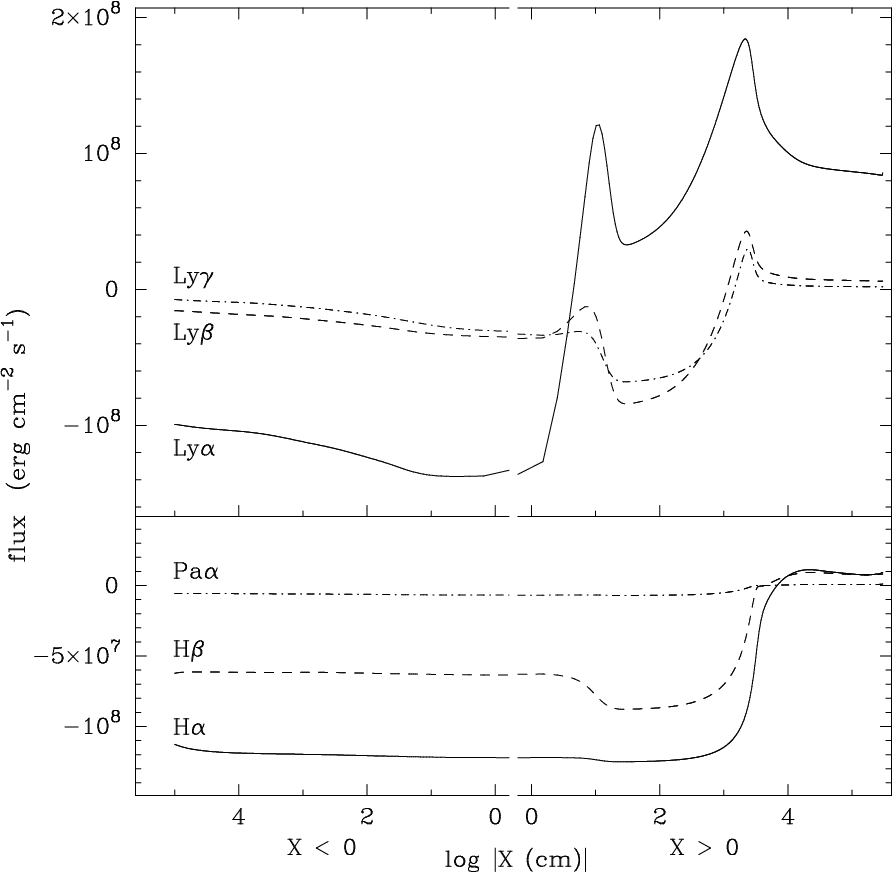}}
\caption{The radiation flux integrated over spectral line frequency
intervals as a function of space coordinate $X$ for the shock wave model
with $\rho_1=10^{-10}~\gcc$, $T_1=6000\K$ and $U_1=60~\kms$.}
\label{frline}
\end{figure}

In Fig.~\ref{frline} we show the plots of the radiation flux integrated
within frequency intervals of the \Lya, \Lyb, \Lyg\ (upper panel),
\Ha, \Hb\ and \Pa\ (lower panel) lines.
Comparing with plots of $n_2/\nH$ and $\xH$ displayed in Fig.~\ref{flux}
we can conclude that the spectral line radiation is produced mostly in
the layers of hydrogen recombination.
Another interesting feature is that the upstream and downstream \Lya\
radiation fluxes are nearly equal in absolute value,
whereas the radiation flux in
Balmer lines emerges mostly upstream.

\section{Radiative losses of the shock wave}

\begin{figure}
\resizebox{\hsize}{!}{\includegraphics{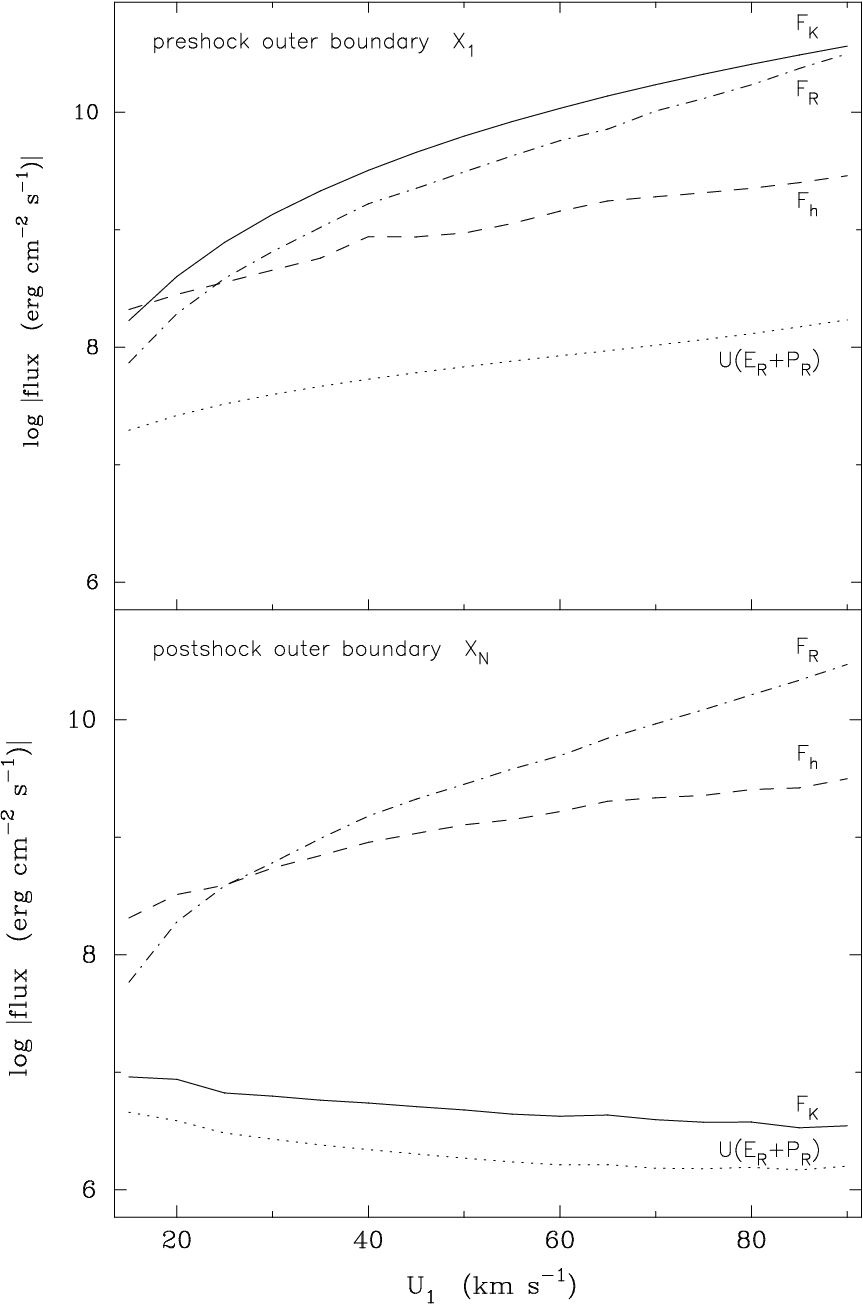}}
\caption{The flux of kinetic energy $\FK$ (solid lines),
the flux of enthalpy $\Fh$ (dashed lines),
the radiation flux $\FR$ (dot--dashed lines),
and the flux $U(\ER+\PR)$ (dotted lines)
at the preshock (upper panel) and postshock (lower panel) outer boundaries
of the slab as a function of upstream velocity $U_1$
in the models with $\rho_1=10^{-10}~\gcc$ and $T_1=6000\K$.}
\label{flux1n}
\end{figure}

In order to estimate the irreversible loss occurring in the shock wave energy
it is instructive to compare the values of each term of the left--hand
side of relation (\ref{CC2}) at both boundaries of the slab.
In Fig.~\ref{flux1n} these quantities are shown for shock wave models
with $\rho_1=10^{-10}~\gcc$, $T_1=6000\K$ and $20~\kms\le U_1\le 80~\kms$.
As is seen by comparing the upper and lower panels representing
the preshock and postshock outer boundaries, respectively, we see that 
most of the kinetic energy of the gas flow is irreversibly lost,
since the ratio of the fluxes of kinetic energy at both boundaries of the
slab ranges within $10^2\lesssim\FKa/\FKn\lesssim 7\cdot 10^3$.
At the same time, the increase of the enthalpy flux after the passage
of the parcel of gas through the slab does not exceed 20\%.
Thus, most of the energy of the shock wave is lost due to radiation.

\begin{figure}
\resizebox{\hsize}{!}{\includegraphics{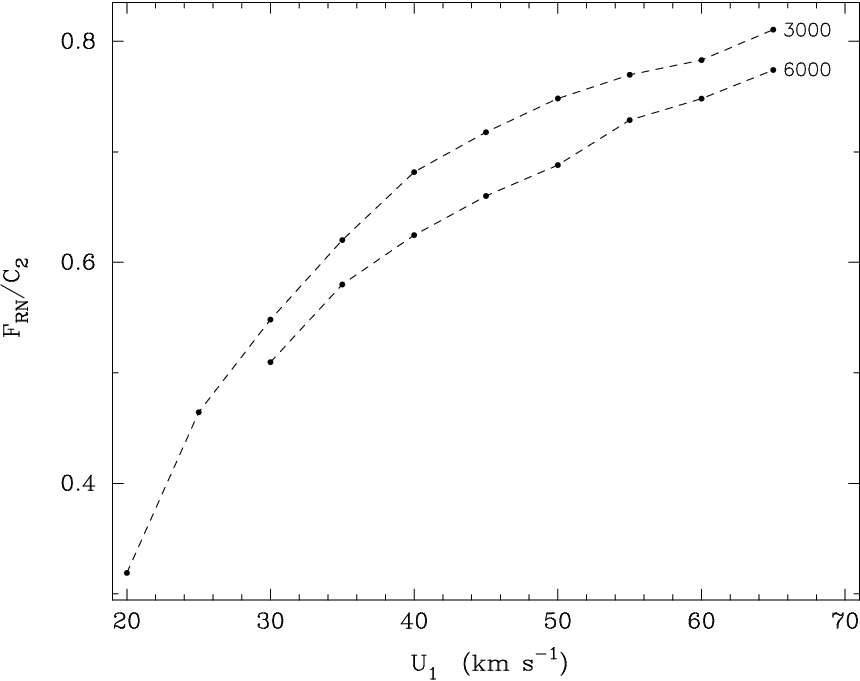}}
\caption{The ratio of the downstream radiation flux to the
total energy flux  $\FRn/\CC_2$ at the postshock outer boundary as a
function of upstream velocity $U_1$ for shock wave models with
$\rho_1=10^{-10}~\gcc$.
Each curve is labeled with $T_1$.}
\label{frcc}
\end{figure}

The role of radiation in irreversible energy losses of the shock wave
can be evaluated from comparison of the radiation flux emerging from the
boundary of the slab with the total energy flux $\CC_2$.
In Fig.~\ref{frcc} the plots of the ratio $\FRn/\CC_2$ at the
postshock outer boundary are shown for two sequences of shock wave models with
$T_1=3000\K$ and $T_1=6000\K$.
As is seen, the radiative losses increase very rapidly with
increasing upstream velocity $U_1$, though there is also the
dependence on the preshock gas temperature $T_1$.

\section{Conclusion}

In this work we improved the treatment of radiation transfer and considered
the structure of shock waves propagating through partially ionized
hydrogen gas with densities from $10^{-12}~\gcc$ to $10^{-9}~\gcc$.
Our calculations showed that while the unpertubed hydrogen gas is partially 
ionized, the preshock gas temperature affects the shock wave structure,
mostly due to the preshock hydrogen ionization because the postshock
equilibration of
translational degrees of freedom is very sensitive to the number density
of hydrogen ions.
In general, a higher temperature in the preshock gas will lead to
larger postshock ionization and to a stronger radiation field
produced by the shock wave.
At the same time, the increase in the preshock gas temperature is accompanied
by the slight decrease of the postshock gas compression.

The preshock gas density $\rho_1$ affects the shock wave structure mostly
due to the fact that the preshock flux of kinetic energy as well as the
radiation flux emerging from the shock wave are proportional to $\rho_1$.
Therefore, in the gas with lower density, the shock wave produces
weaker radiation flux and less compression of the postshock gas.
Moreover, at lower gas density, the width of the shock wave increases.
The width of the radiative precursor increases due to the smaller
absorption in the Lyman continuum, whereas the width of the postshock
relaxation zone increases because of the proportionality of the rates
of relaxation processes to the gas density.

The increase in the shock wave velocity leads to higher rates of
relaxation processes in the compressed gas
behind the discontinuous jump and to higher postshock ionization.
While the maximum ionization degree of the postshock gas is less than unity,
the width of the shock wave decreases with increasing shock wave velocity.
However, when the postshock gas is fully ionized, the increase in the
shock wave velocity leads to the extension of the postshock zone of
hydrogen ionization and, therefore, to an increase in the shock wave width.

Strong radiative cooling of the postshock gas leads to compression
ratios as high as $\rho/\rho_1\sim 10^2$.
Such large increases in the postshock gas density obviously can favour the
condensation of dust grains in outer atmospheres of radially pulsating
late--type giants.

The isothermal approximation undoubtedly overestimates the radiative losses
of the shock wave energy by at least a factor of two for shocks with an
upstream velocity of $U_1\lesssim 30~\kms$ and can provide good accuracy
at upstream velocities of $U_1 > 80~\kms$, that is at Mach numbers
$M_1\gtrsim 10$.

Our calculations demonstrated that the monochromatic radiation flux at
frequencies of hydrogen lines significantly exceeds the flux of the
background continuum, so that our models reproduce the
emission spectrum observed in radially pulsating late--type supergiants.
Moreover, the fraction of the energy transported in hydrogen lines is
not negligible and can be as large as one or two percent of the total
radiation flux because of the strong broadening of the line profiles behind
the discontinuous jump.

In this paper, we did not discuss the details of the emission line spectra
because current results are insufficient for comparison with
observations.
The change in the velocity of the gas during its passage
through the shock wave must inevitably influence the line profiles.
Calculations of shock wave structure based on the solution of the
transfer equation which takes into account effects of the gas flow
velocity gradient were beyond the scope of the present work
and will be presented in the forthcoming paper.

\begin{acknowledgements}
The work of YAF was done in part under the auspices of the
Aix-Marseille~I University and CNRS.
YAF acknowledges also the support from the Russian Foundation for Basic
Research (grant 98--02--16734).
\end{acknowledgements}


\end{document}